\documentclass[twocolumn]{aastex6}

\usepackage{amsmath}

\begin{document}


\title{Deep-HiTS: Rotation Invariant Convolutional Neural Network for
  Transient Detection}



\author{Guillermo Cabrera-Vives\altaffilmark{1,2,3,5}, Ignacio
  Reyes\altaffilmark{4,1,5}, Francisco F\"orster\altaffilmark{2,1},
  Pablo A. Est\'evez\altaffilmark{4,1} and Juan-Carlos
  Maureira\altaffilmark{2}\\Email: gcabrera@dim.uchile.cl}

\altaffiltext{1}{Millennium Institute of Astrophysics, Chile}
\altaffiltext{2}{Center for Mathematical Modeling, Universidad de Chile, Chile}
\altaffiltext{3}{AURA Observatory in Chile}
\altaffiltext{4}{Department of Electrical Engineering, Universidad de Chile, Chile}
\altaffiltext{5}{These authors contributed equally to this work}
%
%
%




\begin{abstract}
We introduce Deep-HiTS, a rotation invariant convolutional neural
network (CNN) model for classifying images of transients candidates
into artifacts or real sources for the High cadence Transient Survey
(HiTS). CNNs have the advantage of learning the features automatically
from the data while achieving high performance. We compare our CNN
model against a feature engineering approach using random forests
(RF). We show that our CNN significantly outperforms the RF model
reducing the error by almost half. Furthermore, for a fixed number of
approximately 2,000 allowed false transient candidates per night we
are able to reduce the miss-classified real transients by
approximately 1/5. To the best of our knowledge, this is the first
time CNNs have been used to detect astronomical transient events. Our
approach will be very useful when processing images from next
generation instruments such as the Large Synoptic Survey Telescope
(LSST). We have made all our code and data available to the community
for the sake of allowing further developments and comparisons at
https://github.com/guille-c/Deep-HiTS\footnote{Deep-HiTS is licensed
  under the terms of the GNU General Public License v3.0.}.

\end{abstract}




\section{Introduction} \label{sec:intro}

As in many other fields, large scale survey telescopes are already
generating more data than humans are able to process, posing the need
for new data-processing tools. Some examples are the Sloan Digital
Sky Survey \cite[SDSS,][]{York2000}, the Panoramic Survey Telescope
and Rapid Response System \citep[Pan-STARRS,][]{Hodapp2004}, the
All-Sky Automated Survey for Supernovae
\citep[ASAS-SN,][]{Shappee2014}, and the Dark Energy Survey
\citep[DES,][]{DES2016} among others. Furthermore, in the near future
we expect to have the Large Synoptic Survey Telescope
\citep[LSST,][]{LSST2009} operative, which will scan the whole
southern sky every couple of days producing terabytes of data per
night. These instruments allow us to explore not only the space
domain, but also the time domain opening new opportunities for
understanding our universe.

The High cadence Transient Survey \citep[HiTS,][]{Forster2016}, is
aimed at detecting and following up optical transients with
characteristic timescales from hours to days. The primary goal of HiTS
is to detect supernovae (SNe) during their earliest hours of
explosion. HiTS uses the Dark Energy Camera
\citep[DECam,][]{Flaugher2015} mounted at the prime focus of the
Victor M. Blanco 4 m telescope on Cerro Tololo near La Serena, Chile.
The HiTS pipeline detects transients using difference images: a
template image is subtracted from the science image taken at the time
of observation and point sources are located within this difference
image. A custom made pipeline does the astrometry, PSF matching,
candidate extraction, and candidate classification into real or fake
transients. Our previous approach uses random forests
\citep[RF,][]{Breiman2001} over a set of handmade features for
candidate classification. Similar approaches have been used in the
past by different groups including \cite{Romano2006},
\cite{Bloom2012}, \cite{Brink2013}, and \cite{Goldstein2015}. The
feature engineering approach usually requires an important amount of
work done by the scientist in order to create representative features
for the problem at hand.

An alternative approach is to learn the features from the data
itself. Convolutional neural networks \citep[CNNs,][]{Fukushima1980}
are a type of artificial neural network \citep[ANN,][]{McCulloch1943}
that have recently gained interest in the machine learning community.
CNNs have achieved remarkable results in image processing challenges,
outperforming previous approaches \citep[e.g.][]{Krizhevsky2012,
  Razavian2014, Szegedy2014}. Though CNNs have been applied to a
variety of image processing problems, they have only recently caught
the attention of the Astronomy community mainly thanks to the Galaxy
Zoo Challenge. \cite{Dieleman2015} won the contest using a rotation
invariant CNN approach. This approach was latter extended to data from
the Hubble Space Telescope in \cite{Huertas2015}.

In this paper we introduce Deep-HiTS, a framework for
transient detection based on rotational invariant deep convolutional
neural networks, and present the advantages and improved results
obtained using this framework. We start by explaining the HiTS
difference image data in Section \ref{sec:data}. In Section
\ref{sec:CNN} we explain the basics of ANN and CNN, as well as
describing our model architecture. We then describe our experiments
and show how the proposed CNN model significantly outperforms the feature
engineering + RF approach in Section \ref{sec:experiments}. We finally
summarise this work and conclude in Section \ref{sec:conclusions}.


\section{Data} \label{sec:data}

The HiTS pipeline finds candidates by using the difference of a
science and a template image and it produces image stamps of $21\times
21$ pixels for each candidate. In this paper we use data from the 2013
run, in which we observed 40 fields in the $u$ band every
approximately 2 hours during 4 consecutive nights.  For the
classification model we use four stamps: template image, science
image, difference image, and signal-to-noise ratio (SNR) difference
image (the difference image divided by the estimated local noise).

For the purpose of training the classification model, we created a
data-set of real non-transients (negatives hereafter) and
simulated transients (positives hereafter). Negatives were produced by
running the first steps of the HiTS pipeline including the astrometry,
PSF matching, and candidate extraction over observed
  images. We evaluate the proposed model on the next step, which is
candidate classification into non-transient or
transient. Negatives are artifacts caused mostly by
  statistical fluctuations of the background, inaccurate astrometry,
  and bad CCD columns. This produced 802,087 candidates which we
labeled as negatives\footnote{We are aware this data-set contains some
  point like transients not present in the reference image, but we
  conservatively estimate they will be less than a
  0.2\%.}. Notice these negatives were produced by running the
  pipeline over real data, hence no simulation process is used for
  obtaining them.

\begin{figure*}[ht!]
\centering
\begin{tabular}{cc}
NEGATIVES & POSITIVES\\
\includegraphics[width=0.45\textwidth]{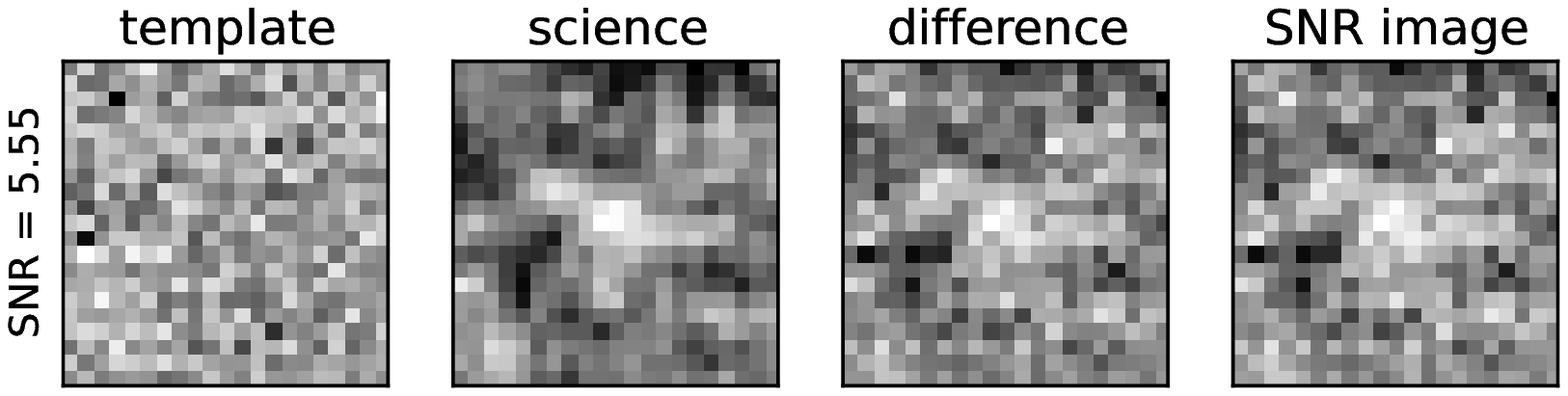} &
\includegraphics[width=0.45\textwidth]{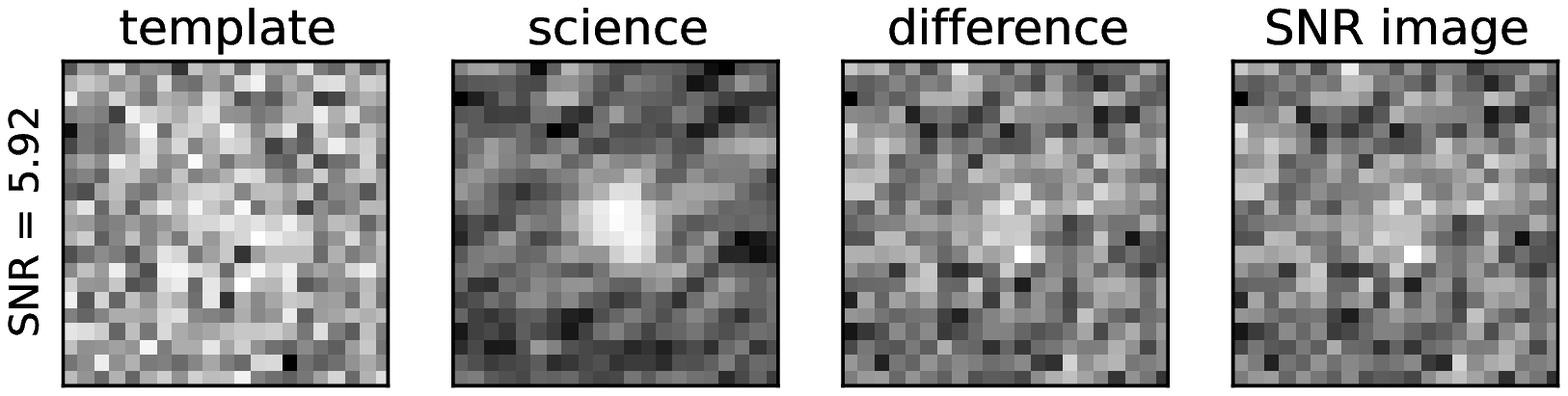}\\ 
\includegraphics[width=0.45\textwidth]{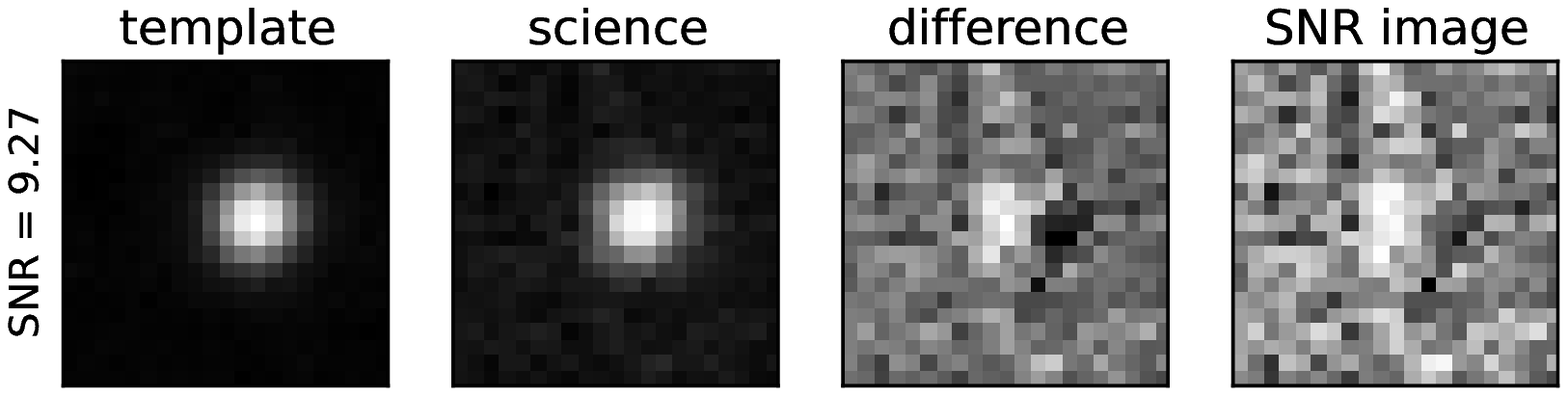} &
\includegraphics[width=0.45\textwidth]{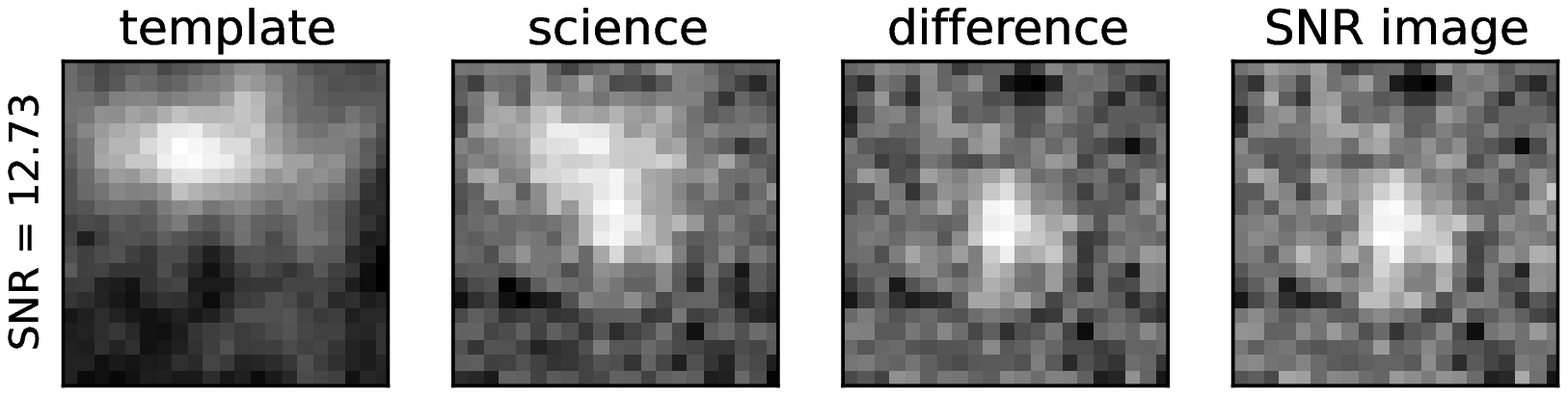}\\ 
\includegraphics[width=0.45\textwidth]{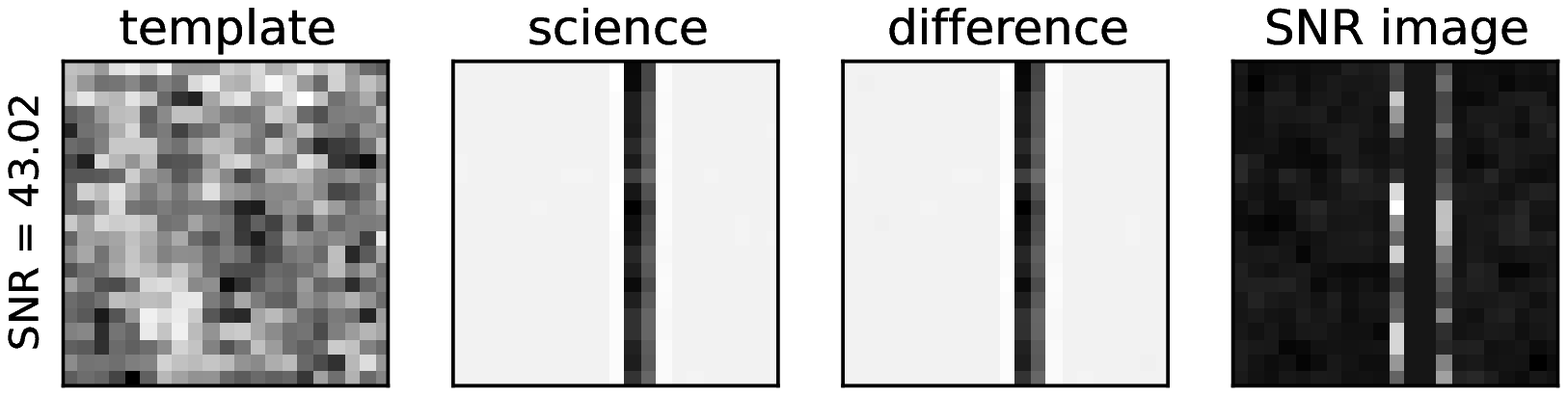} & 
\includegraphics[width=0.45\textwidth]{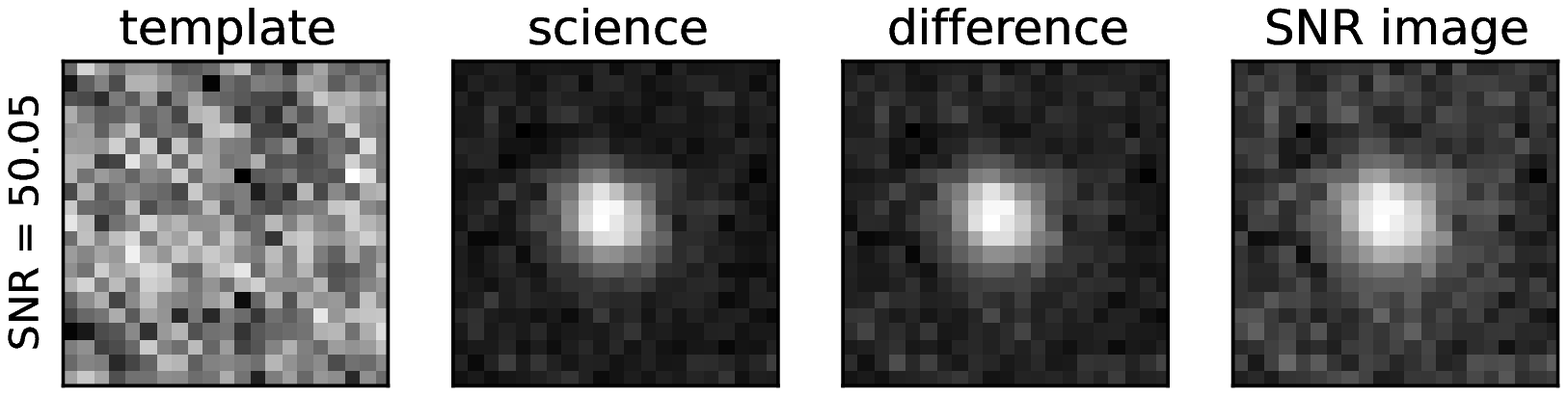}\\
\end{tabular}
\caption{Examples of artifacts (negatives) and simulated transients
  (positives) at different signal-to-noise ratios. From top to bottom
  artifacts are caused by statistical fluctuations of the background,
  inaccurate astrometry, and bad CCD column.}
\label{fig:candidates}
\end{figure*}

Positives were generated by selecting stamps of real PSF-like sources
and placing them at a different location at the same epoch and in the
same CCD they were observed. By doing this we simulated positives
using real point like sources, hence reproducing how transients would
look like. Point sources are obtained by first selecting sextractor
\citep{Bertin1996} isolated sources which are present in both frames
with a consistent position and flux, and whose: size is within 2
median absolute deviations from the median size of the sources of the
image, total flux is positive, and minimum pixel value including sky
emission is greater than zero. We also filtered out those sources that
show a sextractor flux inconsistent with its optimal photometry flux
\citep{Naylor1998}, which we found to be a good discriminator between
stars and galaxies. Positives are then added scaled to mimic the SNR
distribution of the candidates found in the image (fitted power law
distribution). Images with these simulated transients are run through
the first steps of the HiTS pipeline (astrometry, PSF matching and
candidate extraction) in order to get stamps of positives. As
explained above, non-simulated detected candidates are taken as
negatives. In order to keep a balanced data-set, we used the same
number of positives as negatives. Figure \ref{fig:candidates} shows
examples of negative and positive candidates and their respective
stamps.


\section{Convolutional Neural Network Model} \label{sec:CNN}



\subsection{Artificial neural networks}

CNNs are a type of artificial neural network \citep[ANN, see][and
  references therein]{Zhang2000}.
  The basic unit of an ANN is called neuron. Each neuron receives a
  vector as input and performs a dot product operation between that
  vector and a set of parameters called \emph{weights}. The resulting
  value plus a bias goes into a non-linear activation function,
  usually a sigmoid. Neurons are grouped in layers, each one with its
  own weights. In a multilayer perceptron, layers are stacked one
  after the other. The output of layer $n$, $\mathbf{x}_n$, depends on
  the output of layer $n-1$, $\mathbf{x}_{n-1}$, and is given by
\begin{equation}
  \mathbf{x}_n= f(\mathbf{W}_n\cdot \mathbf{x}_{n-1} + b_n),
  \label{eq:mlp}
\end{equation}
where $\mathbf{W}_n$ is the weight matrix, $b_n$ is the bias, and $f$
is the activation function. These layers are called fully-connected
layers.


ANNs are usually trained using an algorithm called error
backpropagation, which adjust the weights of the network in an
iterative way proportional to the gradient descent of the error with
respect to the weights.
In this work neural networks are trained using stochastic gradient
descent \citep[SGD,][]{LeCun1998}, where the gradient of the error
with respect to the weights is estimated at each iteration using a
small part of the data-set called mini-batch. That means that for
every iteration the gradient is estimated as the average instantaneous
gradients for a small group of samples (usually between 10 and 500),
which provides a good trade-off in terms of gradient estimation
stability and computing time.  

The learning rule for SGD is given
by
\begin{equation}
  \theta_{t+1} = \theta_{t} - \eta \nabla_{\theta}\mathcal{L},
  \label{eq_sgd}
\end{equation}
where $\theta$ represents a parameter of the ANN (such as
weights or biases), $\mathcal{L}$ is the loss function (e.g. mean squared
error) and $\eta$ is the learning rate.  The learning rate controls
the size of the steps that SGD makes in each iteration. Reducing the
learning rate during training
allows having a good compromise between exploration at the beginning
and exploitation (local fine tuning) in the final part of the
process.

\subsection{Convolutional neural networks}

Convolutional neural networks (CNNs) usually contain two
types of layers: convolutional and pooling layers. Convolutional
layers perform a convolution operation between the input of the layer
and a set of weights called \emph{kernel} or \emph{filter}. We extend
Equation \ref{eq:mlp} by considering the input of the $n$
convolutional layer to be a stack of $K$ arrays $\mathbf{x}_{k, n-1}$
($k=1,\ldots,K$), the outputs as $L$ arrays $\mathbf{x}_{l, n}$
($l=1,\ldots,L$) and the filters as matrices $\mathbf{W}_{k,l,
  n}$. The output of layer $n$ is then obtained as
\begin{equation}
\mathbf{x}_{n, l} = f\left(\sum_{k=1}^{K}\mathbf{W}_{k,l,n}\ast\mathbf{x}_{k, n-1} + b_{l, n}\right),
\end{equation}
where $\ast$ is the convolution operator, and $b_{l, n}$ are the
biases of layer $n$. In order to add some flexibility on the output size
of a layer, zeros can be appended in the input layer's borders. This
is called \emph{zero-padding}.

Pooling layers return a subsampled version of the input data. In the
pooling layer the data is divided into small windows and a single
representative value is chosen for each window, e.g. the maximum
(max-pooling) or the average (mean-pooling). A common design for CNNs
consist on a mix of convolutional and pooling layers at a first stage
followed by dense fully-connected layers.

Though sigmoids are usually chosen as activation functions, they are
not the only choice.  Rectified linear units \citep[ReLU,
][]{Nair2010} are activation functions that have gained popularity
in the last years because they can achieve fast training and good
performance. The output of a ReLU is the maximum between the input and
zero
\begin{equation}
f(x) = \max(0, x).
\end{equation}
Leaky ReLUs 
are variants where the negative inputs are not
set to zero but instead they are multiplied by a small number
(e.g. $0.01$) to preserve gradient propagation,
\begin{equation}
  f(x)  = 
  \begin{cases}
    x     & \mbox{if } x > 0, \\
    0.01x & \mbox{otherwise}.
  \end{cases}
\end{equation}

In order to reduce overfitting, a usual technique is 
\emph{dropout} \citep{Srivastava2014}, which consists in turning off
random neurons during training time with a probability $p$,
usually chosen as $0.5$. The goal of dropout is to prevent
coadaptation of the outputs, hence neurons do not depend on other
neurons being present in the network. In order to preserve the scale
of the total input, the remaining neurons need to be rescaled by
$1/(1-p)$. Dropout is activated only during training time. When
evaluating models, all neurons become active.

\subsection{Deep-HiTS architecture}

\begin{figure*}[ht!] 
\includegraphics[width=0.9\textwidth]{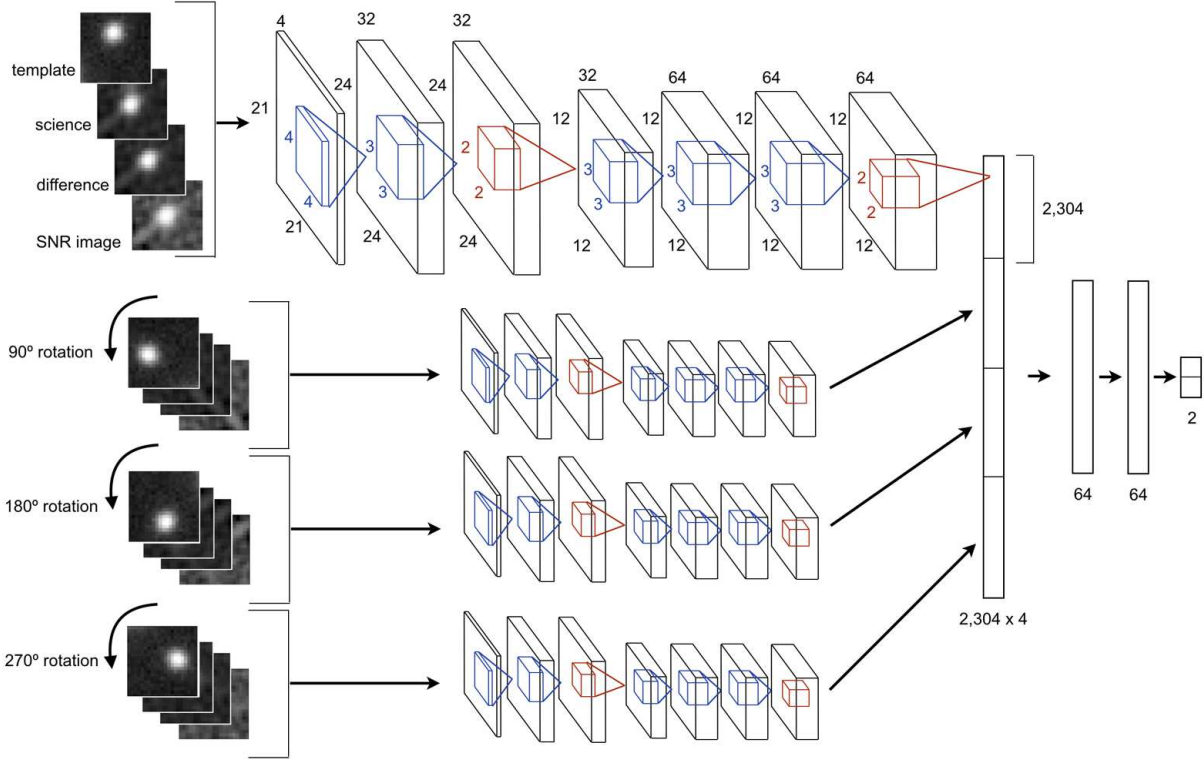} 
\caption{Architecture of the proposed rotational invariant CNN
  model. Blue boxes represent convolutional layers and red boxes
  represent max-pooling layers. The stacked images of each candidate
  are presented to the network rotated four times in order to exploit
  rotational symmetry. The calculated feature maps of each rotation
  are flattened and stacked creating a $2,304\times 4$ vector that is
  fed to a fully connected network whose outputs are the probabilities
  of being an artefact and a real transient. }\label{fig:architecture}
\end{figure*}

Preliminary results were presented in a conference paper by
\cite{Cabrera2016} where a very basic CNN model was proposed achieving
a performance comparable to the RF model. Here we extend these results
by using a much deeper architecture and introducing rotational
invariance as well as leaky ReLUs and dropout, significantly
outperforming the RF model.
Following \cite{Dieleman2015} we introduce rotation invariance by
applying the convolutional filters to various rotated versions of the
image, thus exploiting rotational symmetry present in transients
images.
Figure \ref{fig:architecture} shows the proposed rotation invariant
convolutional neural network architecture.
 The template, science, difference, and SNR difference stamps are
presented to the network as a stacked array of size
$21\times 21 \times 4$. The convolutional architecture composed by the
convolutional and pooling layers yields 2,304 features. Rotation
invariance is incorporated by presenting every stamp to the
convolutional architecture rotated four times, hence producing
$2,304\times 4$ features. These 9,216 features are fed to three fully
connected layers that return the probabilities of being an artefact
and a real transient.

After trying more than 50 different architectures and training
  strategies\footnote{Most important improvements were
    achieved by adding more layers, number of parameters per layer,
    different activation functions, different training strategies
    (learning rates, dropout), and adding rotation invariance.} the
best performance was obtained by the architecture shown in
Figure \ref{fig:architecture}.  Our convolutional architecture is
composed of two convolutional layers with filters of size $4\times 4$
and $3\times 3$ respectively, a $2\times 2$ max-pooling layer, three
$3 \times 3$ convolutional layers and a $2\times 2$ max-pooling
layer. Zero-padding was used in order to set the sizes of each layer
to the ones shown in Figure \ref{fig:architecture} ($24\times 24$ for
the first two and $12\times 12$ for the last four). All convolutional
layers use ReLUs. The concatenated output of all rotations are fed to
three stacked fully connected layers: two ReLU layers of 64 units
each, and a final logistic regression layer that outputs two
probabilities: fake and real transient.

\section{Experiments} \label{sec:experiments}

\begin{figure}[ht!]
\plotone{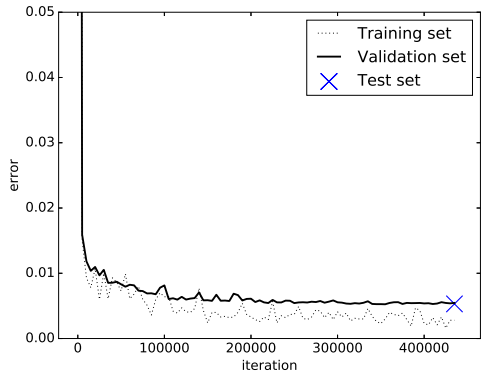}
\caption{Learning curve (evolution of the error during the training
  process). The training set error is calculated over the last 10,000
  candidates that were used for training. The test set error is
  calculated at the end of training. Validation error does not raise
  at the end of training, indicating that the model is not
  overfitted.}
\label{fig:learning_history}
\end{figure}

We used the candidates data-set described in Section \ref{sec:data} to
assess the performance of the proposed CNN model and compare it
against our previous random forest model. We split the data into
1,220,000 candidates for training, 100,000 candidates for validation,
and 100,000 for testing. The CNN model is trained using SGD with
mini-batches of 50 examples from the training set and 0.5
dropout. Learning rate was reduced to half every 100,000 iterations,
starting with an initial value of 0.04. The validation data-set was
used for measuring the performance of the model at training time and
decide when to stop training. We assumed the model converged when
after feeding 100,000 candidates the zero-one loss (fraction of
misclassifications) does not go lower than a 99\% of the previous
loss.
Once the model is trained, we use the test set (data unseen by the
model) to calculate performance metrics. Figure
\ref{fig:learning_history} shows the zero-one loss of the model as a
function of the number of iterations for the training and validation
sets, and for the test set after training. Due to insufficient memory
to store the whole training set, the training set curve is calculated
over the last 10,000 candidates used for training at each iteration,
while the validation curve shows the loss for the whole 100,000
candidates in the validation set. We programmed our code using Theano
\citep{Theano2016} and trained our model on a tesla K20 graphics
processor unit (GPU). The model converged after 455,000 iterations and
took roughly 37 hours to train achieving an error of 0.525\% over the
validation data-set and 0.531\% over the test set for this particular
experiment.

We compare the performance of the proposed CNN model against our
previous RF model \citep[implemented using
scikit-learn][]{Pedregosa2011}. As stated above, we consider
transients to be positives ($P$) and non-transients to be negatives
($N$). We use the following metrics to assess the performance of our
models:
\begin{align}
\text{accuracy} = & \frac{TP + TN}{TP + FN + TN + FP},\\
\text{precision} = & \frac{TP}{TP + FP},\\
\text{recall} = & \frac{TP}{TP + FN},\\
\text{f1-score} = & 2\frac{\mathrm{precision\cdot recall}}{\mathrm{precision+ recall}} 
= \frac{2 TP}{2 TP + FP + FN},
\end{align}
where $TP$ stands for true positives (number of positives correctly
classified), $TN$ stands for true negatives (number of negatives
correctly classified), $FP$ stands for false positives (number of
negatives incorrectly classified as positives by the model), and $FN$
stands for false negatives (number of positives incorrectly classified
as negatives by the model).
In order to make a fair comparison, we trained and tested the RF model
with the same number of candidates (1,220,000 for training and 100,000
for testing). For both CNN and RF models, six train-validation-test
splits were done using stratified shuffle split (i.e. maintaining
positives and negatives balanced for all sets). Table \ref{tab:RF_CNN}
shows the mean accuracy, precision, recall, and f1-score obtained with
the RF and CNN models and their respective standard deviations. All
metrics show that the CNN model outperforms the RF model. To assess
the statistical significance of these results, we performed a Welch's
hypothesis test \citep{Welch1947} obtaining a probability of less than
a $10^{-7}$ that these results were obtained by chance.

\begin{deluxetable*}{c|CCCC}
\label{tab:RF_CNN}
\tablecaption{Comparison of  RF and CNN models. CNN shows consistent
  better results than RF in terms of accuracy, precision, recall and
  f1-score. Results are statistically significant based on the
  calculated Welch's T-test p-values, being all of them lower than
  $10^{-7}$. Accuracy results show that the error of CNN model is
  almost half the error obtained with the RF model.}
\tablehead{\colhead{model} & \colhead{accuracy} & \colhead{precision}
  & \colhead{recall} & \colhead{f1-score}}
\startdata
RF & 
98.96 \pm 0.03\,\%&
98.74 \pm 0.06\,\%&
99.18 \pm 0.02\,\%&
98.96 \pm 0.03\,\%\\
CNN &         \,
99.45 \pm 0.03\,\%&
99.30 \pm 0.07\,\%&
99.59 \pm 0.07\,\%&
99.45 \pm 0.03\,\%\\
\hline
Welch's T-test p-value &
4.7\times 10^{-11}&
2.8\times 10^{-08}&
9.7\times 10^{-08}&
4.5\times 10^{-11}\\
\enddata
\end{deluxetable*}

\begin{figure*}[ht!]
\centering
\begin{tabular}{cc}
\includegraphics[width=0.45\textwidth]{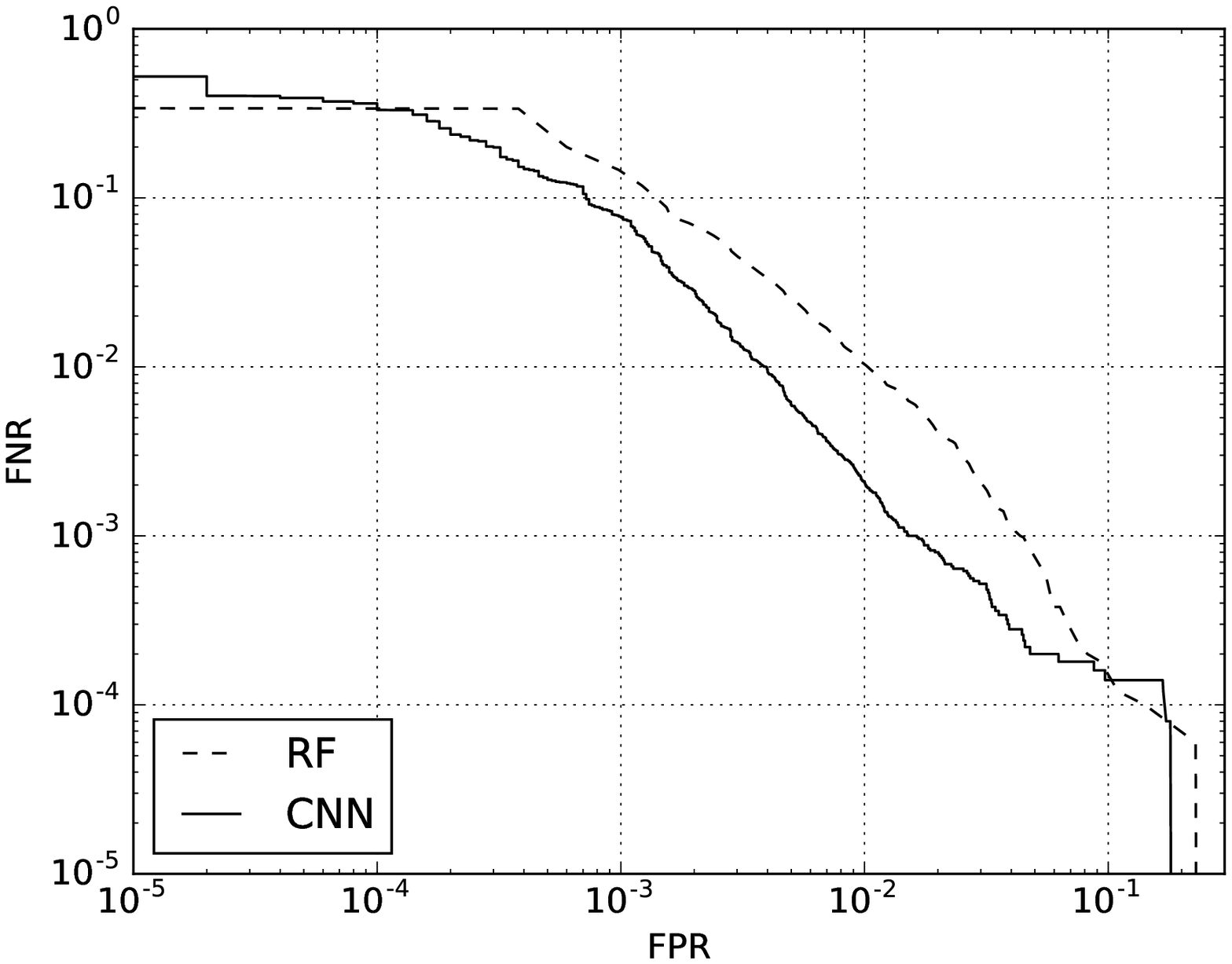} &
\includegraphics[width=0.45\textwidth]{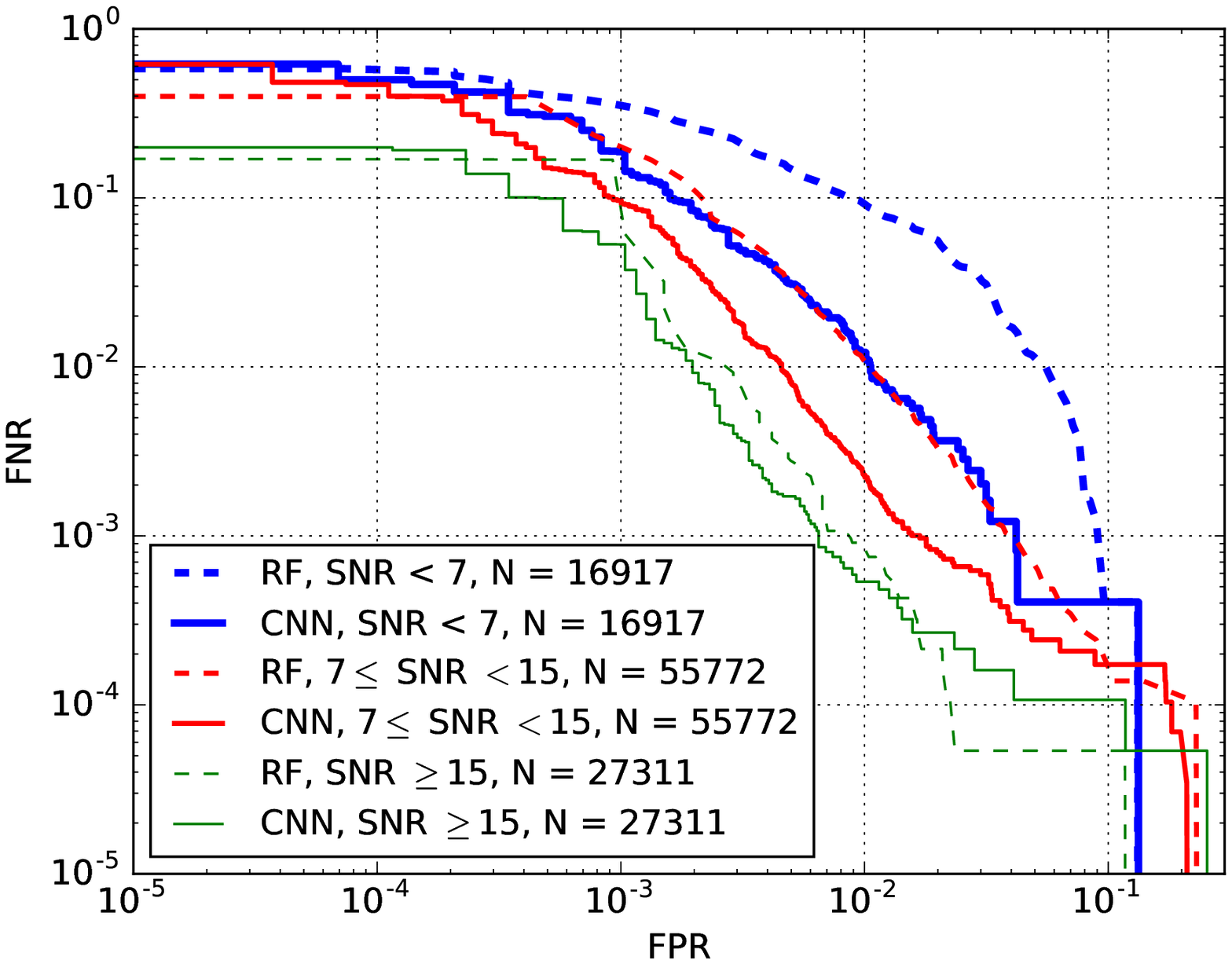}\\ 
(a)&(b)
\end{tabular}
\caption{Detection error tradeoff (DET) curve comparing a Random
  Forest model against the proposed CNN. DET curve plots False
  Negative Rate (FNR) versus False Positive
  Rate (FPR). A curve closer to the lower bottom indicates better
  performance. (a) Overall DET curves. 
  (b) DET curves for subsets of candidates binned according to their
  signal-to-noise ratio (SNR), where $N$ indicates the number of
  candidates per bin.  The proposed CNN model achieves a better
  performance than the RF model for a FPR between $10^{-4}$ and
  $10^{-1}$, where the HiTS pipeline operates.  The biggest improvement
  occurs for the lowest SNR candidates.}
  \label{fig:DET}
\end{figure*}

Figure \ref{fig:DET} shows the detection error tradeoff (DET) curves
of the RF and CNN models. The DET curve plots the false negative rate
(FNR) versus the false positive rate (FPR) as we vary the probability
threshold used to determine positives and negatives, where
\begin{equation}
\text{FPR} = \frac{FP}{N}, \qquad\text{FNR} = \frac{FN}{P}.
\end{equation}
Figure \ref{fig:DET}a shows the overall DET curves for the RF and CNN
models, while Figure \ref{fig:DET}b shows the DET curves for different
SNR ranges. A better model would obtain smaller FPR and FNR, hence
moving the curve to the bottom left of the plot.  It can be seen again
that the proposed CNN model achieves better performance than our
previous RF model. The DET curve is also useful for evaluating the
trade-off between false positives and false negatives.  In practice,
when observing in real time, most candidates will be negatives. All
negatives on our data-set are real candidates taken from the HiTS
pipeline: 802,087 obtained during 4 consecutive nights,
i.e. approximately 200,000 negatives per night. The number of
estimated real transients is considerably smaller than that.  As an
example, consider 
as user-defined operation point getting 2,000 false positives per
night (FPR $\sim 10 ^{-2}$).  Figure \ref{fig:DET} tells us that by
using the proposed CNN we will get a FNR $\sim 2\times10^{-3}$ which
is much better in comparison to a FNR $\sim 10^{-2}$ when using the RF
model. This means 5 times less real transients would be missed by
using the CNN model. On Figure \ref{fig:DET}b we can see that this
improvement is mostly achieved by the proposed CNN being able to
correctly classify low SNR sources.

\begin{figure}[ht!]
\includegraphics[width=0.45\textwidth]{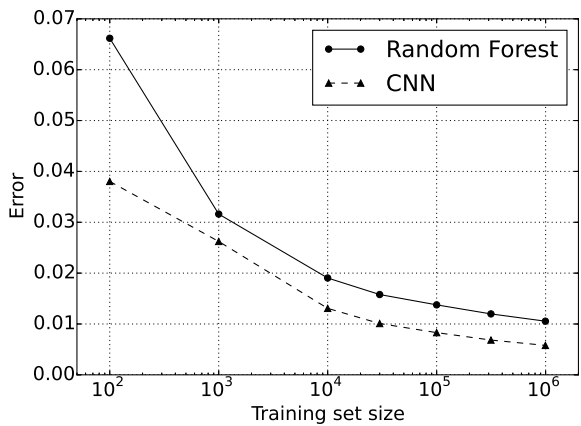}
\caption{Error over a validation data-set of 100,000 candidates in
  terms of the size of the training set for the Random Forest and
  Convolutional Neural Network models.}
  \label{fig:acc_vs_N}
\end{figure}

In Figure \ref{fig:acc_vs_N} we explore the performance of the
  RF and CNN models in terms of the size of the training set. For this
  purpose we saved 100,000 candidates for validation and trained each
  model with training sets of different sizes. It can be seen that the
  CNN model outperforms the RF model independently of the size of the
  training set used. Furthermore, by using a RF model trained over
  $10^6$ candidates we obtain similar performance as using a CNN model
  trained over $3\times 10^4$. 

\begin{figure}[ht!]
\includegraphics[width=0.45\textwidth]{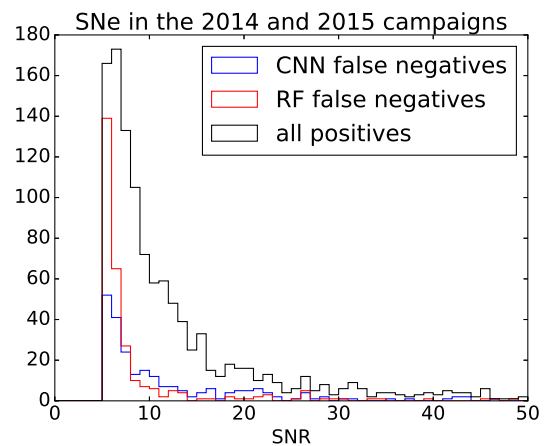}
\caption{Comparison of RF and CNN models over real SNe from the 2014 and
  2015 campaigns.}
  \label{fig:2014-2015}
\end{figure}

In order to test our framework over real data, we applied the RF and
CNN models over SNe found in the 2014 and 2015 campaigns. For our 2013
campaign we observed in the $u$ band filter, while in the 2014 and
2015 campaigns we observed in the $g$ band filter. We used stamps from
SNe at all epochs with a SNR over 5. During observation time HiTS
found 47 SNe in the 2014 campaign and 110 in the 2015 campaign, giving
a total of 1166 candidates (each SN has candidates at different
epochs). Using the user-defined FPR of $10 ^{-2}$ described above, the
RF model correctly classified 866 of these positive candidates
(FNR = 0.257), while Deep-HiTS correctly classified 921 of
them (FNR = 0.210).
Notice that though we trained our models on $u$ band images, we are
still able to classify candidates in the $g$ band filter. 
Figure \ref{fig:2014-2015} shows the
number of positive candidates incorrectly classified for the SNe
present in the 2014 and 2015 campaigns for the RF and CNN models in
terms of the SNR. It can be seen that for a SNR below 8 the CNN model
outperforms the RF model. These are the most important candidates when
detecting transients in real time, as we hope to detect them at an
early stage. For a SNR over 8 the RF model better recovers the true
positives, which suggests the use of a hybrid model between RF and
CNN. We will expand on this idea in future work.

As we tested our model on real data from a different HiTS campaign,
and also using a band filter different from the one used for training,
we did not achieve the low FNR shown in Figure \ref{fig:DET}. However
our CNN model is able to recover more true positives from images in
the g-band filter than the RF-based model given the user-defined
operation point described above.  We currently do not have a robust
way of measuring the false positives automatically in the 2014 and
2015 data-sets as the new observational setup yields a significantly
larger population of unknown asteroids which would need to be removed
from the set of negatives. Further research is needed in order to
assess the best way of transferring our CNN model between different
data-sets in terms of recovering positives and negatives.


\section{Conclusions}\label{sec:conclusions}

We introduced Deep-HiTS, a rotation invariant deep convolutional
neural network (CNN) for detecting transients in survey
images. Deep-HiTS discriminates between real and fake transients in
images from the High cadence Transient Survey (HiTS), a survey that
aims to find transients with characteristic timescales from hours to
days using the Dark Energy Camera (DECam). We compared the proposed
CNN model against our previous approach based on a random forest (RF)
classifier trained over features designed by hand. Deep-HiTS not only
outperforms the RF approach, but it also has the advantage of learning
suitable features for the problem automatically from the data.  The
proposed classification model was able to improve the overall accuracy
of the HiTS pipeline from $98.96 \pm 0.03\,\%$ to
$99.45 \pm 0.03\,\%$. In practice, by using Deep-HiTS we can reduce
the number of missed transients to approximately 1/5 when accepting
around 2,000 transients per night. The use of deep learning models
over next generation surveys, such as the LSST may have a great impact
on detecting and classifying the unknown unknowns of our
universe. Deep-HiTS is open source and available at GitHub
(\url{https://github.com/guille-c/Deep-HiTS}), and the version used in this
paper is archived on Zenodo (\url{https://doi.org/10.5281/zenodo.190760}).


\acknowledgments 

We gratefully acknowledge financial support from CONICYT-Chile
through its FONDECYT postdoctoral grant number
3160747; CONICYT-PCHA through its national M.Sc. scholarship 2016 number
22162464; CONICYT through the Fondecyt
Initiation into Research project No. 11130228; CONICYT-Chile through grant Fondecyt 1140816; CONICYT-Chile and NSF through the Programme of
International Cooperation project DPI201400090; CONICYT through the
infrastructure Quimal project No. 140003; Basal Project PFB--03; the
Ministry of Economy, Development, and Tourism's Millennium Science
Initiative through grant IC120009, awarded to The Millennium Institute
of Astrophysics (MAS).
Powered@NLHPC: this research was partially supported by the
supercomputing infrastructure of the NLHPC (ECM-02). We acknowledge
the usage of the Belka GPU cluster (Fondequip EQM 140101). 
This project used data obtained with the Dark Energy Camera (DECam),
which was constructed by the Dark Energy Survey (DES)
collaboration. Funding for the DES Projects has been provided by the
U.S. Department of Energy, the U.S. National Science Foundation, the
Ministry of Science and Education of Spain, the Science and Technology
Facilities Council of the United Kingdom, the Higher Education Funding
Council for England, the National Center for Supercomputing
Applications at the University of Illinois at Urbana-Champaign, the
Kavli Institute of Cosmological Physics at the University of Chicago,
Center for Cosmology and Astro-Particle Physics at the Ohio State
University, the Mitchell Institute for Fundamental Physics and
Astronomy at Texas A\&M University, Financiadora de Estudos e Projetos,
Fundação Carlos Chagas Filho de Amparo, Financiadora de Estudos e
Projetos, Fundação Carlos Chagas Filho de Amparo à Pesquisa do Estado
do Rio de Janeiro, Conselho Nacional de Desenvolvimento Científico e
Tecnológico and the Ministério da Ciência, Tecnologia e Inovação, the
Deutsche Forschungsgemeinschaft and the Collaborating Institutions in
the Dark Energy Survey. The Collaborating Institutions are Argonne
National Laboratory, the University of California at Santa Cruz, the
University of Cambridge, Centro de Investigaciones Enérgeticas,
Medioambientales y Tecnológicas–Madrid, the University of Chicago,
University College London, the DES-Brazil Consortium, the University
of Edinburgh, the Eidgenössische Technische Hochschule (ETH) Zürich,
Fermi National Accelerator Laboratory, the University of Illinois at
Urbana-Champaign, the Institut de Ciències de l'Espai (IEEC/CSIC), the
Institut de Física d'Altes Energies, Lawrence Berkeley National
Laboratory, the Ludwig-Maximilians Universität München and the
associated Excellence Cluster Universe, the University of Michigan,
the National Optical Astronomy Observatory, the University of
Nottingham, the Ohio State University, the University of Pennsylvania,
the University of Portsmouth, SLAC National Accelerator Laboratory,
Stanford University, the University of Sussex, and Texas A\&M
University.
All plots were created using matplotlib \citep{Hunter2007}.

\end{document}